\documentclass{ptephy}%%%%where ptephy is the template name

\begin{document}

\title{Application of the nuclear equation of state  \\
obtained by the variational method \\
to core-collapse supernovae}

\author{\name{H.~Togashi}{1,\ast}, \name{M.~Takano}{1,2}, \name{K.~Sumiyoshi}{3}, and \name{K.~Nakazato}{4}}
%%%%%%%%%%% The \name command should be used as \name{Insert author name here}{Insert affiliation number here}
%%%%% Please use \thanks for contributed author details

%%%%%%%%%%% The \affil command should be used as \affil{Insert affiliation number here}{Insert author address here}
\address{\affil{1}{Department of Pure and Applied Physics, 
Graduate School of Advanced Science and Engineering, 
Waseda University, 3-4-1 Okubo Shinjuku-ku, Tokyo 169-8555, Japan}
\affil{2}{Research Institute for Science and Engineering, 
Waseda University, 3-4-1 Okubo Shinjuku-ku, Tokyo 169-8555, Japan}
\affil{3}{Numazu College of Technology, 
Ooka 3600, Numazu, Shizuoka 410-8501, Japan}
\affil{4}{Department of Physics, Faculty of Science and Technology, 
Tokyo University of Science, Yamazaki 2641, Noda, Chiba 278-8510, Japan}
\email{hajime\_togashi@ruri.waseda.jp}}

\begin{abstract}
The equation of state (EOS) for hot asymmetric nuclear matter 
that is constructed with the variational method starting from the Argonne v18 and Urbana IX nuclear forces is applied to spherically symmetric core-collapse supernovae (SNe). 
We first investigate the EOS of isentropic $\beta$-stable SN matter, and find that matter with the variational EOS is more neutron-rich than that with the Shen EOS.  
Using the variational EOS for uniform matter supplemented by the Shen EOS for non-uniform matter at low densities, 
we perform general-relativistic spherically symmetric simulations of core-collapse SNe with and without neutrino transfer, starting from a presupernova model of 15 $M_{\odot}$. 
In the adiabatic simulation without neutrino transfer, the explosion is successful, and the explosion energy with the variational EOS is larger than that with the Shen EOS.  
In the case of the simulation with neutrino transfer, the shock wave stalls and then the explosion fails, as in other spherically symmetric simulations.  
The inner core with the variational EOS is more compact than that with the Shen EOS, due to the relative softness of the variational EOS.  
This implies that the variational EOS is more advantageous for SN explosions than the Shen EOS.  
\end{abstract}

\subjectindex{D41, E26}

\maketitle

\section{Introduction}
In studies of core-collapse supernovae (SNe), 
the equation of state (EOS) of hot dense nuclear matter is one of the essential ingredients, and it has been developed in parallel with the progress of the research on core-collapse SNe.  
In the early stages of these studies, simple,  analytically expressed nuclear EOSs were adopted, 
and the influence of the stiffness parameters included in the EOSs on the SN explosion was discussed \cite{BCK, TS}. 
Hillebrandt and Wolff adopted a more realistic nuclear model: They constructed a nuclear EOS table based on the Skyrme--Hartree--Fock approach \cite{HW}.  
After this study, Lattimer and Swesty (LS) \cite{LS} provided an EOS for SN (SN-EOS) with use of a Skyrme-type interaction for uniform matter and a compressible liquid-drop model for non-uniform matter: 
This EOS has been widely adopted in numerical studies of SNe.  
About ten years later, Shen et al. \cite{Shen1, Shen2, Shen3} constructed a new nuclear EOS based on the relativistic mean field (RMF) theory: 
In their studies, non-uniform nuclear matter is treated in the Thomas--Fermi (TF) approximation.  
The Shen EOS was provided as a table of thermodynamic quantities over an extremely wide range of densities, temperatures and proton fractions, and has been utilized conveniently in complicated numerical simulations of SNe.  
In recent years, the Shen EOS has been extended so as to take into account hyperons  \cite{Ishizuka} and quarks \cite{Nakazato}.  
In addition, RMF Lagrangians different from that adopted in the Shen EOS have been applied to the SN-EOS \cite{Hempel, GShen, Hempel2}.  
Furthermore, the EOS of non-uniform matter has also been improved: 
Instead of the single-nucleus approximation adopted in the LS EOS and the Shen EOS, multiple nuclides are taken into account in Refs. \cite{Hempel, GShen, Hempel2, Furusawa, Mishustin}.  

In these EOSs, as mentioned above, uniform matter is treated with phenomenological nuclear models.  
Since the nuclear EOS is governed by nuclear forces, it is desirable for nuclear EOSs constructed with many-body calculations starting from bare nuclear forces to be directly applied to SNe.  
In fact, nuclear EOSs of this type have been calculated for uniform matter, especially at zero temperature, and have been applied to the study of cold neutron stars (CNSs). 
However, studies of the EOS of hot asymmetric nuclear matter starting from bare nuclear forces are limited.  
It is also noted that, in the case of the Shen EOS, the parameter set of the RMF Lagrangian called TM1\cite{TM1} is determined so that scalar and vector potentials of nuclear matter are close to those obtained by the Dirac--Brueckner--Hartree--Fock calculations \cite{RBHF} with the nuclear potential Bonn A, but the Bonn A potential is not directly adopted in the Shen EOS.  
To our knowledge, there has been no study in which EOSs with microscopic many-body theories starting from bare nuclear forces have been applied directly to SN numerical simulations.  

Toward a more microscopic nuclear EOS, therefore, we are now constructing a new type of SN-EOS based on the realistic nuclear forces using the variational many-body theory.  
As the first step of our project, 
we have performed variational calculations starting from the Argonne v18 (AV18) two-body nuclear potential \cite{AV18} and the Urbana IX (UIX) three-body potential \cite{UIX1, UIX2} 
for hot asymmetric nuclear matter, as reported in Ref.  \cite{TT}.  
In this calculation, we have adopted a relatively simple cluster variational method in order to treat asymmetric nuclear matter, because the more sophisticated Fermi hypernetted chain (FHNC) variational method is hardly applicable to asymmetric nuclear matter.  
In spite of the large difference in the degree of sophistication between these two methods, the obtained (free) energies per nucleon for symmetric nuclear matter and neutron matter in the present EOS are in good agreement with those obtained with the FHNC calculations performed by Akmal et al. (APR) \cite{APR} at zero temperature and by Mukherjee \cite{AM} at finite temperatures.  

We are now at the stage of applying the uniform nuclear EOS based on the cluster variational method to core-collapse SNe; this application is the main aim of the study described in this paper. 
As the first step of our application, we perform general-relativistic, spherically symmetric hydrodynamic numerical simulations of SNe, and compare the results using the variational EOS with those using the Shen EOS as a reference SN-EOS.  
Here, we consider two SN simulations with and without neutrino transfer.  
Without the neutrino transfer, the weak reaction is switched off, and then the electron fraction of each fluid element is fixed during the simulation. 
Namely, the proton fractions of fluid elements are determined by the initial progenitor model, resulting in relatively large proton fractions, and then the dynamics is governed by the nuclear EOS with those large proton fractions.  
Therefore, the simulation without the neutrino transfer is not only suitable as a test of the application of our EOS to SN numerical codes, but also valuable as a study of the properties of the EOS with large proton fractions, though such a case is rather hypothetical.  
Then, as a more realistic case, we perform the SN simulation with the neutrino transfer by taking into account the weak reactions.  
In this case, the electron fraction decreases significantly by neutrino emissions, and then the EOS of neutron-rich nuclear matter plays an important role in the simulation.  

It is well known that, in general, the SN explosion is not successful in spherically symmetric calculations.  
We note that a successful 1D explosion has been reported for a progenitor of a low-mass star with an O--Ne--Mg core in Refs. \cite{Kitaura, Fischer}. 
We consider here the collapse of the iron core of a massive star. 
In this case, 2 or 3D simulations are inevitably required in order to describe convection, 
which is an essential dynamical mode to ensure that fluid elements stay in the heating region longer to gain more energy from the neutrinos emitted from the inside of the core.  
In addition, the standing accretion shock instability mainly promoting dipole oscillation of the shock wave around 
proto-neutron stars (PNSs), which is found in multi-dimensional core-collapse simulations, may be advantageous for stellar explosions (see, e.g., Refs. \cite{SNreview1, SNreview2}). 
Although these multi-dimensional effects cannot be treated in the spherically symmetric simulations, 
they become important at relatively late stages of the SN explosions. 
In the early stages of SNe, the gravitational collapse of the iron core of the progenitor and the subsequent core bounce, where the EOS of dense uniform matter plays essential roles, are well approximated in the spherically symmetric simulations.  
Therefore, as the first step of the application of the variational EOS to SNe, the spherically symmetric hydrodynamic simulation is appropriate.  

In order to perform the SN simulation, we also need a non-uniform EOS: 
The SN-EOS must cover an extremely wide range of densities, temperatures, and proton fractions, in contrast with the case of CNSs.  
We adopt, for simplicity, the non-uniform part of the Shen EOS, and connect it to the variational EOS for uniform matter in the SN simulations.  

In the next section, we review the uniform nuclear EOS adopted in this paper, and discuss some of the thermodynamic quantities of uniform  isentropic SN matter with the variational EOS.  
In Sect.~3, we apply the variational EOS to a spherically symmetric general-relativistic hydrodynamical calculation of a core-collapse SN without the weak reactions.  
In Sect.~4, we apply the variational EOS to a more realistic SN simulation taking into account neutrino transfer.  
A summary and concluding remarks are given in Sect.~5.  

 %%%%%%%%%%%%%%%%%%%
\section{The equation of state of supernova matter with the variational method}
\subsection{Variational calculations of hot asymmetric nuclear matter}
In this section, we briefly review the nuclear EOS that we have constructed in Refs. \cite{TT, K1, K2}, and, based on it, derive an EOS of uniform SN matter.  
We start from the nuclear Hamiltonian composed of the AV18 two-body potential and the UIX three-body potential.  
At zero temperature, we first calculate the expectation value of the Hamiltonian without the three-body force in the two-body cluster approximation to obtain the two-body energy per nucleon $E_2/N$.  
In this calculation, we employ the Jastrow wave function defined by 
\begin{equation}
\mathnormal{\Psi}=\mathrm{Sym}\left[\prod_{i<j}f_{ij}\right]\mathnormal{\Phi}_\mathrm{F},   \label{eq:jwf}
\end{equation}
where $f_{ij}$ is composed of spin--isospin-dependent central, tensor, and spin--orbit correlation functions, which may also depend on the third component of the two-nucleon total isospin so as to treat asymmetric nuclear matter with arbitrary proton fractions $Y_{\mathrm{p}}$.   
$\mathrm{Sym}$[ ] is the symmetrizer with respect to the order of the factors in the products, and $\mathnormal{\Phi}_\mathrm{F}$ is the degenerate Fermi-gas wave function.  
The correlation functions are determined through the minimization of $E_2/N$ under the extended Mayer's condition and the healing distance condition: 
The former is a kind of normalization condition and the latter introduces a healing distance for the correlation functions.  
We assume that the healing distance is proportional to the mean distance between nucleons, and the coefficient is chosen so that the obtained 
$E_2/N$ of symmetric nuclear matter and neutron matter are close to those obtained with the FHNC method by APR.  

The energy caused by the three-body force, $E_3/N$, is constructed based on the expectation value of the three-body Hamiltonian with the degenerate Fermi-gas wave function.  
Here, with use of the uncertainty of the three-body force, the strength of the repulsive and two-pion-exchange components of the UIX potential are modified and a correction term, 
which is an explicit function of the nucleon number density $n_{\mathrm{B}}$ and the proton fraction $Y_{\mathrm{p}}$, is added somewhat phenomenologically.  
The values of the four parameters included in $E_3/N$ are chosen so that the total energy per nucleon $E/N =E_2/N +E_3/N$ reproduces the empirical saturation density $n_0$, saturation energy $E_0/N$, incompressibility $K$, and symmetry energy $E_{\mathrm{sym}}$.  
Furthermore, these parameters in $E_3/N$ are fine-tuned so that the Thomas--Fermi calculations of isolated atomic nuclei with the present $E/N$ reproduce the gross features of their experimental masses and radii, as reported in Ref. \cite{K2}.  
%%%%%%%%%%%%%%%%%%%%%%%%%%%%%%%%%%%%20131206 The explicit values 
%20140107 Explicitly, the obtained values for the present EOS 
The obtained values for the present EOS 
%%%%%%%%%%%%%%%%%%%%%%%%%%%%%%%%%%%%%%%20140106 
are $n_0 = 0.16$ fm$^{-3}$, $E_0/N = -16.09$ MeV, $K = 245$ MeV, and $E_{\mathrm{sym}} = 30.0$ MeV \cite{TT}.
We note that the uncertainties in the empirical values of $K$ and $E_{\mathrm{sym}}$ are relatively large.  
The above-mentioned value of $K$ is consistent with $K = 240 \pm 10$ MeV deduced from the measurements of the isoscalar giant monopole resonance in medium-mass nuclei, as reported in Refs. \cite{Colo, Piekarewicz}.  
Similarly, the value of $E_{\mathrm{sym}}$ lies in the range $28 \lesssim E_{\mathrm{sym}} \lesssim 34$ MeV derived from the combination of constraints from experimental and CNS observational studies, as reported in Ref. \cite{SHF}
and references therein.  

The obtained $E/N$ for symmetric nuclear matter and neutron matter are in good agreement with the results of APR.  
In addition, the density dependence of the symmetry energy is consistent with the experimental data on heavy ion collisions \cite{expS1, expS2}. 
Furthermore, we have constructed the EOS of non-uniform matter at zero temperature with the TF calculation using the Wigner--Seitz approximation, and applied it to CNS crusts \cite{K2}. 
Then, we have calculated the structure of CNSs.  
The maximum mass of CNSs is $2.22$ $M_\odot$, and the obtained mass--radius relation is consistent with the observational data analyzed by Steiner et al.\cite{Steiner}.    
It is noted that the maximum mass with the present EOS is consistent with the recently observed masses of heavy CNSs \cite{2Msolar1, 2Msolar2}.  

At finite temperatures, we have extended the variational method proposed by Schmidt and Pandharipande \cite{SP1, SP2} so as to calculate the free energy per nucleon $F/N$ of asymmetric nuclear matter.  
In this method, we express $F/N$ at temperature $T$ as 
\begin{equation}
\frac{F}{N} = \frac{E_{\mathrm{0T}}}{N}-T\frac{S_0}{N}.    \label{eq:FE}
\end{equation}
The approximate internal energy per nucleon $E_{\mathrm{0T}}/N$ is the sum of the two-body energy $E_{\mathrm{2T}}/N$ and the three-body energy $E_{\mathrm{3T}}/N$.
$E_{\mathrm{2T}}/N$ is constructed by extending the two-body energy at zero temperature, $E_2/N$, by replacing the nucleon occupation probabilities in $\mathnormal{\Phi}_\mathrm{F}$ composing the Jastrow wave function Eq. (\ref{eq:jwf}) by the average occupation probabilities $n_i(k)$ $(i=\mathrm{p}, \mathrm{n})$ for protons and neutrons, respectively.  
The explicit expression of $n_i(k)$ is as follows: 
\begin{equation}
n_i(k) = \Bigg\{1+\exp \left[\frac{\varepsilon_i(k)-\mu_{0i}}{k_{\mathrm{B}}T}\right]\Bigg\}^{-1}.    \label{eq:nk}
\end{equation}
Here, the quasi-nucleon energies $\varepsilon_i(k)=\hbar^2k^2/2m_i^*$ are specified by the effective masses $m_i^*$.  
For simplicity, $E_{\mathrm{3T}}/N$ is assumed to be the same as $E_3/N$ at zero temperature.  
The approximate entropy per nucleon $S_0/N$ is also expressed with $n_i(k)$ as 
\begin{equation}
\frac{S_0}{N} = -\sum_{i=\mathrm{p,n}} \frac{k_{\mathrm{B}}}{\pi^2 n_{\mathrm{B}}}\int_0^{\infty}
\Bigl\{\left[1-n_i(k)\right]\ln\left[1-n_i(k)\right]+n_i(k)\ln n_i(k)\Bigl\}k^2dk.
\end{equation}
Then, $F/N$ is minimized with respect to $m^\ast_{\mathrm{p}}$ and $m^\ast_{\mathrm{n}}$.  

The obtained $F/N$ of symmetric nuclear matter and neutron matter are consistent with those obtained with the FHNC method \cite{AM}, and thermodynamic self-consistency is confirmed, i.e., $E_{\mathrm{0T}}/N$ coincides with the internal energy per nucleon derived from the obtained $F/N$ through the thermodynamic relation.  
The obtained critical temperature of symmetric nuclear matter is about 18 MeV. 
 
\subsection{The equation of state of isentropic neutrino-trapped supernova matter}
Using the variational EOS reported above, we investigate some of the thermodynamic quantities of uniform, neutrino-trapped SN matter, as compared with those with the Shen EOS.  
Namely, we consider charge-neutral and $\beta$-stable mixtures of nucleons, leptons, and photons, with fixed lepton number fractions per nucleon $Y_{\mathrm{l}}$. 
As the leptons, we consider $\mathrm{e}^-$, $\mathrm{e}^+$, $\nu_{\mathrm{e}}$, and $\bar{\nu}_{\mathrm{e}}$, and treat them as relativistic non-interacting Fermi gases.  

\begin{figure}[t]
  \centering
  \includegraphics[width=10.0cm]{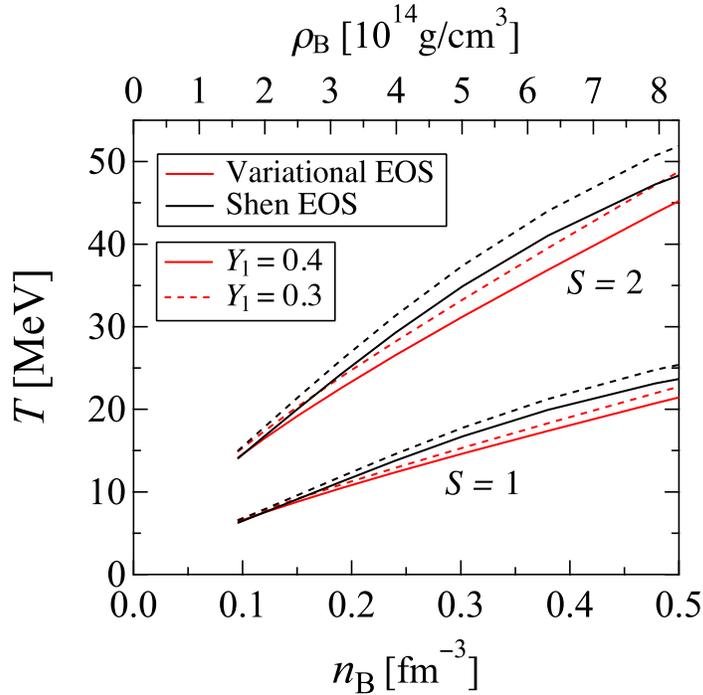}
  \caption{The temperatures of the isentropic SN matter for $Y_{\mathrm{l}}$ = 0.3 and 0.4 at $S$ = 1 and 2 as functions of the baryon number density $n_{\mathrm{B}}$. }
\label{fig:PNStemp}
\end{figure}

Figure \ref{fig:PNStemp} shows the temperature $T$ of isentropic SN matter as a function of the baryon number density $n_{\mathrm{B}}$ for $Y_{\mathrm{l}}$ = 0.3 and 0.4 at the entropy per baryon $S$ = 1 and 2 in units of the Boltzmann constant $k_{\mathrm{B}}$.  
Also shown in this figure are the results with the Shen EOS.  
It can be seen that $T$ with the variational EOS at fixed $S$ and $Y_{\mathrm{l}}$ are close to those with the Shen EOS, though the results with the variational EOS are slightly lower in this density region.  
In Fig. \ref{fig:PNStemp}, the baryon mass density $\rho_{\mathrm{B}}$ corresponding to $n_{\mathrm{B}}$ is also shown.  
In this paper, we employ the definition $\rho_{\mathrm{B}} = n_{\mathrm{B}}m_{\mathrm{u}}$ with $m_{\mathrm{u}}$ being the atomic mass unit, 
by following the description of the Shen EOS \cite{Shen1, Shen2, Shen3}.  

\begin{figure}[t]
  \centering
  \includegraphics[width=14.0cm]{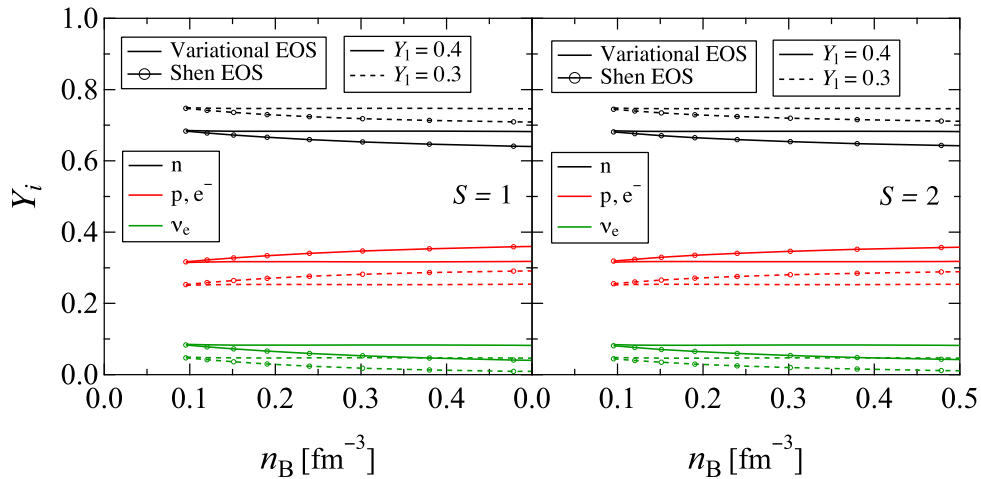}
  \caption{The fractions of particles $Y_i$ of isentropic SN matter for $Y_{\mathrm{l}}$ = 0.3 and 0.4 at $S$ = 1 (left panel) and 2 (right panel) 
  as functions of the baryon number density $n_{\mathrm{B}}$. }
  \label{fig:SNMYi}
\end{figure}

The fractions of neutrons, protons, electrons, and electron-neutrinos $Y_i$ ($i$ = n, p, e$^-$, and $\nu_{\mathrm{e}}$) are shown in Fig. \ref{fig:SNMYi} as functions of the baryon number density $n_{\mathrm{B}}$ for isentropic SN matter at $S$ = 1 (left panel) and 2 (right panel). 
The neutron fraction $Y_{\mathrm{n}}$ is the largest, and the proton fraction $Y_{\mathrm{p}}$ coincides with the electron fraction $Y_{\mathrm{e}}$ due to charge neutrality.  
There is no appreciable difference between the fractions with $S$ = 1 and 2.
In contrast, the increase in $Y_{\mathrm{l}}$ affects the particle fractions considerably: 
In the case of $Y_{\mathrm{l}}$ = 0.4, $Y_{\mathrm{p}}$ and $Y_{{\nu}_{\mathrm{e}}}$ 
become larger than in the case of $Y_{\mathrm{l}}$ = 0.3, while $Y_{\mathrm{n}}$ becomes smaller.  
The corresponding particle fractions in the case of the Shen EOS are also shown in this figure.  
It can be seen that $Y_{\mathrm{n}}$ and $Y_{{\nu}_{\mathrm{e}}}$ with the variational EOS are larger than those with the Shen EOS, while $Y_{\mathrm{p}}$ with the present EOS is smaller.  
This is because the symmetry energy of the variational EOS is smaller than that of the Shen EOS in this density region. 
In fact, $E_{\mathrm{sym}}=30.0$ MeV for the variational EOS at the saturation density, while $E_{\mathrm{sym}}=36.9$ MeV for the Shen EOS. 

\begin{figure}[t]
  \centering
  \includegraphics[width=14.0cm]{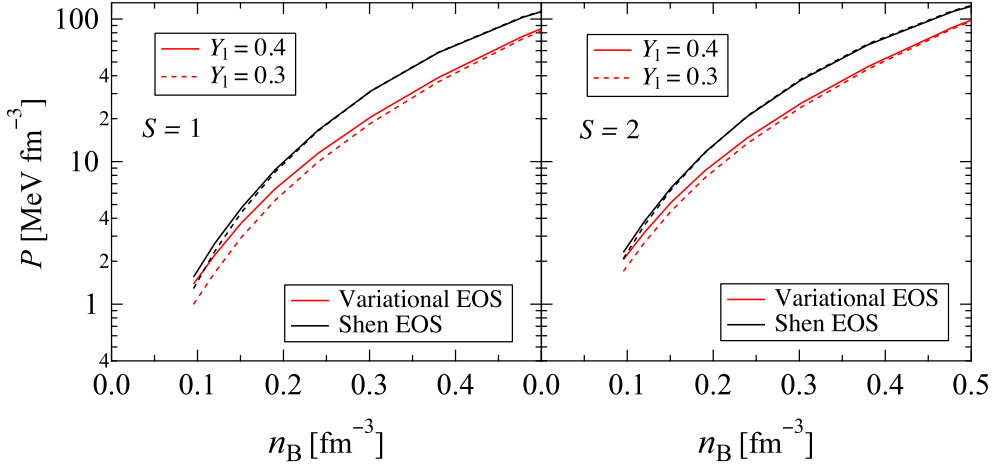}
  \caption{The pressures of the isentropic SN matter for $Y_{\mathrm{l}}$ = 0.3 and 0.4 at $S$ = 1 (left panel) and 2 (right panel)  as functions of the baryon number density $n_{\mathrm{B}}$.
The pressures of the Shen EOS are also shown. }
\label{fig:eos}
\end{figure}

The pressure $P$ of SN matter is shown in Fig. \ref{fig:eos}.  
It is seen that  $P$ increases with $S$ and $Y_{\mathrm{l}}$.  
At the densities $n_{\mathrm{B}} \gtrsim 0.25$ fm$^{-3}$, the contribution from the nucleon pressure to total $P$ is dominant, while at lower densities, the contribution from the leptons and photons dominates.  
Since the central density of the collapsed core of an SN at the bounce is $n_{\mathrm{B}} \sim 0.32$ fm$^{-3}$, as we will see in the following sections, the pressure originating from nucleons is essential at the bounce of the core.  
Comparing the present results with those with the Shen EOS, 
we see that $P$ with the variational EOS at a given set of $S$ and $Y_{\mathrm{l}}$ is smaller than the corresponding value with the Shen EOS.  

It is noted that, at least in the density region shown in these figures, the causality condition is satisfied. 
The critical baryon mass density from which the violation of the causality starts is $\rho_{\mathrm{c}} = 1.5\times10^{15}$g/cm$^{3}$, corresponding to the critical number density $n_{\mathrm{c}} = 0.93$ fm$^{-3}$, for $Y_{\mathrm{l}}$ = 0.3 and $S=1$. 
This value is insensitive to the values of $Y_{\mathrm{l}}$ and $S$.   
On the other hand, the density relevant to SNe is $\rho_{\mathrm{B}} \lesssim 5.2\times10^{14}$g/cm$^{3}$ ($n_{\mathrm{B}} \lesssim  0.32$ fm$^{-3}$), 
which is sufficiently lower than the critical density $\rho_{\mathrm{c}}$.  
Therefore, the present EOS is safely applicable to SN simulations.  

%%%%%%%%%%%%%%%%%%%%%%%%%%%%%%%%%%%%%%%%%%%%%%%%%%%%%%%%%%%%%%%%%%%%%%%%%%%%%%%%
\section{Application to spherical adiabatic core-collapse supernovae}
In this section, we apply the present variational EOS to hydrodynamical numerical simulations of the core-collapse SN.  
We perform fully general-relativistic, spherically symmetric hydrodynamical calculations using the numerical code originally designed by Yamada \cite{SY1}.  
In this code the time evolution of the general-relativistic fluid elements is solved implicitly so that 
long-term calculations with a large time step are possible.  
This code was used in Ref. \cite{KS3} to perform SN simulations with the Shen EOS for various progenitor models in order to study the EOS of dense matter in dynamical situations, in addition to the study of the progenitor dependence of the SN hydrodynamics.  
Since one of our main purposes is to achieve the first application of the variational EOS to the SN hydrodynamic calculations, we choose a typical presupernova model, i.e., the 15 $M_\odot$ model calculated by Woosley and Weaver \cite{WW}.  
In this case, the mass of the iron core is 1.32 $M_\odot$.  
In the hydrodynamical simulations, following the study of Ref. \cite{KS3}, we use the part of the central core that includes the iron core with a density down to about $10^6$g/cm$^3$.  
The mass of the central core treated in this simulation is 1.56 $M_\odot$ \cite{KS3}.  
Details of the structure of the central core, which is used as the initial condition, are reported in Ref. \cite{KS3}.  

As the simplest case of the SN simulations, we perform an adiabatic hydrodynamic calculation.  
In this case, the weak interaction is switched off, and then the electron fraction $Y_{\mathrm{e}}$ of each fluid element does not change throughout the simulation.  
Since neutrinos are absent in the initial progenitor model, they do not appear throughout the simulation, and then there is no energy flow due to neutrino transfer in the SN matter, i.e., the evolution of the SN matter is adiabatic.  
Since  $Y_{\mathrm{e}}$ of each fluid element is maintained at the initial value, which is about 0.4--0.5, the dynamics of the SN is governed by the nuclear EOS with relatively large values of the proton fraction $Y_{\mathrm{p}}$: 
We examine the effect of the variational EOS with such $Y_{\mathrm{p}}$ on the SN hydrodynamics.  
In contrast, when the neutrino transfer is taken into account, the SN matter becomes more neutron-rich.  
In this case, we examine the effect of the EOS with relatively small values of $Y_{\mathrm{p}}$.  
The latter case including the neutrino transfer is reported in the next section.  

In order to apply the present variational EOS to SN simulations, 
we first prepare a table of the free energies of hot asymmetric nuclear matter for the grid points of 
baryon mass densities $\rho_{\mathrm{B}}$, temperatures $T$, and proton fractions $Y_{\mathrm{p}}$ as shown in Table \ref{table1}.  
\begin {table}[b]
\caption{The ranges of the temperature $T$, proton fraction $Y_{\mathrm{p}}$, and baryon mass density $\rho_{\mathrm{B}}$ in the table of the variational EOS. 
In the table, which gives the number of the grid point in the mesh of $T$, the "$+1$" represents the case at $T=0$ MeV.}
\label{table1}
\begin {center}
\begin{tabular}{ccccc} \hline
   Parameter & Minimum & Maximum & Mesh & Number \\ \hline
   $\mathrm{log}_{10} (T)$ [MeV] & $-1.00$ & $2.60$ & $0.04$ & $91+1$   \\  
   $Y_{\mathrm{p}} $ & $0.00$ & $0.65$ & $0.01$ & $66$   \\  
    $\mathrm{log}_{10} (\rho_{\mathrm{B}})$ [g/cm$^3$] & $14.0$ & $16.0$ & $0.10$ & $21$   \\    \hline
\end{tabular}
\end {center}
\end{table}
These grid points are chosen to be the same as those used with the new Shen EOS \cite{Shen3} at high densities $\rho_{\mathrm{B}} \ge 10^{14}$ g/cm$^3$.  
We also prepare other thermodynamic quantities such as the internal energies, entropies, and chemical potentials at the same grid points.  
In order to perform SN simulations, it is necessary to use a non-uniform EOS as well as a uniform EOS:
The SN-EOS must cover an extremely wide range of densities ($10^5$--$10^{15}$g/cm$^3$).  
Since we focus on the influence of the present variational EOS of uniform matter on SN simulations, we supplement the non-uniform EOS at low densities by the Shen EOS.  
Namely, we adopt the variational EOS at $\rho_{\mathrm{B}} \geq 10^{14.3}$g/cm$^3$ and the Shen EOS at $\rho_{\mathrm{B}} \le 10^{14}$ g/cm$^3$; we connect these EOSs in the density region $10^{14}$ g/cm$^3$ $\le \rho_{\mathrm{B}} \le 10^{14.3}$ g/cm$^3$ so that each thermodynamic quantity of isentropic matter including electrons, positrons, and photons at a fixed $Y_{\mathrm{p}}$ vary with respect to $n_{\mathrm{B}}$ smoothly.  
Since, in the initial condition, the central density of the core is about 10$^{10}$g/cm$^3$, the structure of the central core is completely governed by the Shen EOS at that time.  
This initial state is marginally unstable to gravitational collapse, and then it starts collapsing when we start the numerical simulation.  

\begin{figure}[t]
  \centering
  \includegraphics[width=10.0cm]{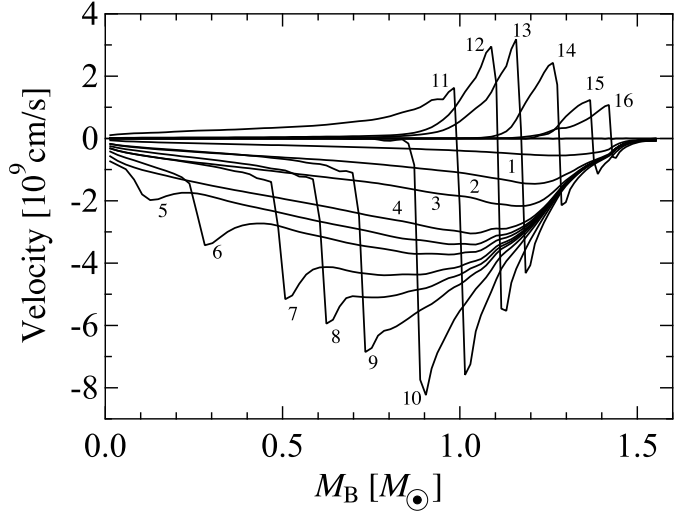}
  \caption{The velocity profiles at selected times as functions of the baryon mass coordinate $M_{\mathrm{B}}$.  
  The numbers denote the time sequence as explained in the text.}
\label{fig:velocity}
\end{figure}

The evolution of the velocity profile of the iron core calculated with the SN simulation is shown in Fig. \ref{fig:velocity}.  
Here, the velocity is given as a function of the baryon mass coordinate $M_{\mathrm{B}}$ in units of $M_{\odot}$, which represents the (time-dependent) radius by the baryon mass lying inside; 
for a more precise definition, see the Appendix of Ref.  \cite{KS3}.  
The curves in this figure correspond to the velocity profiles at selected times, and are numbered for convenience.  
The numbers (1)--(4) represent the times when the central density reaches $10^{11}$, $10^{12}$, $10^{13}$, and $10^{14}$ g/cm$^3$ during the collapse, respectively.  
The number (10) corresponds to the time of the bounce, and we set the time $t_{\mathrm{pb}}$ to zero at the bounce.  
The explicit values of $t_{\mathrm{pb}}$ are shown in Table \ref{table2}.
\begin {table}[b]
\caption{Times corresponding to the numbers in Fig. \ref{fig:velocity}.}
\label{table2}
\begin {center}
\begin{tabular}{ccccccccccccccccc} \hline
    Number & $1$ & $2$ & $3$ & $4$ & $5$ & $6$ & $7$ & $8$ \\
    Time $t_{\mathrm{pb}}$ [ms] & $-41.4$ & $-6.18$ & $-1.84$ & $-0.58$ & $-0.40$ & $-0.30$ & $-0.20$ & $-0.15$   \\  \hline
    Number   & $9$ & $10$ & $11$ & $12$ & $13$ & $14$ & $15$ & $16$ \\
    Time $t_{\mathrm{pb}}$ [ms]   & $-0.10$ & $0.00$ & $0.15$ & $0.50$ & $1.00$ & $5.00$ & $25.0$ & $50.0$  \\  \hline
\end{tabular}
\end {center}
\end{table}

The velocity profiles shown in Fig. \ref{fig:velocity} behave in a way similar to the case of the Shen EOS \cite{KS3}.  
In the first stage (1--4), the matter falls toward the center of the core, and the corresponding velocity profile consists of an inner self-similar collapse and an outer free fall.   
As the density of the matter approaches saturation density (5--9), the matter becomes stiffer, and then the falling materials decelerate to create an outgoing pressure wave.  
When the decelerated fluid elements bounce to attain positive velocities (10), a shock wave is created.  
Then, the shock wave propagates toward the surface of the core (11--15).  
This result implies that the variational EOS constructed in this paper can be successfully applied to sophisticated core-collapse SN simulations.  

\begin{figure}[t]
  \centering
  \includegraphics[width=12.0cm]{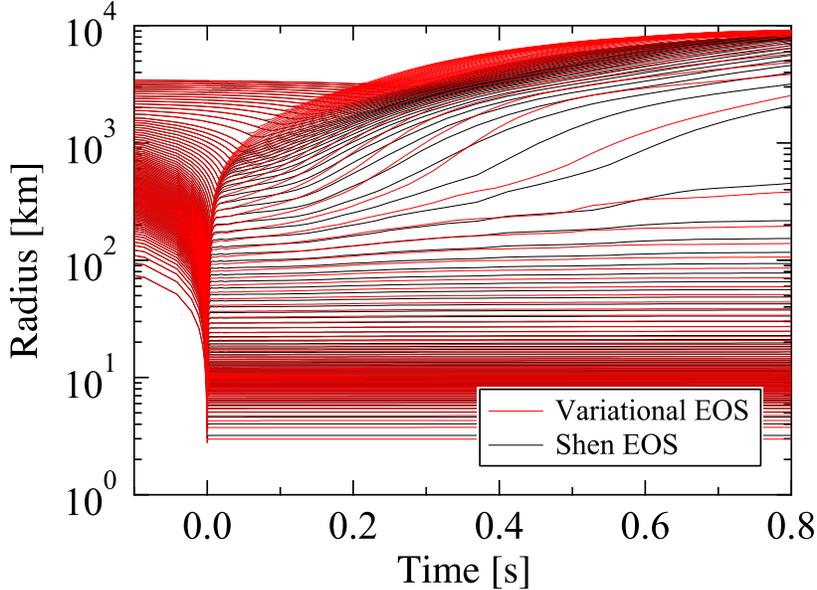}
  \caption{The trajectories of the mass mesh in the radius of the core of a 15 $M_{\odot}$ star as functions of time with our SN-EOS. 
  Also shown are those with the Shen EOS. }
\label{fig:radius}
\end{figure}

Figure \ref{fig:radius} shows the trajectories of the mass mesh in the radius of the central core.  
The red curves are with the present EOS and the black ones are with the Shen EOS.  
Here, as in the case given above, the times at the bounce are chosen to be zero in both cases.  
This figure shows that, before the bounce, the trajectory of each fluid element falling toward the center of the iron core with the present EOS is completely the same as that with the Shen EOS, because the matter is not yet uniform.  
When the density of the matter becomes close to the nuclear saturation density around the time of the bounce, the trajectories with the present EOS deviate from those with the Shen EOS.  
In fact, at the bounce, it is seen in this figure that the lowest curves, corresponding to the innermost grid points, 
fall to smaller radii with the present EOS than those with the Shen EOS.  
This implies that the present EOS is softer than the Shen EOS in this adiabatic case (relatively large values of $Y_{\mathrm{p}}$).  

After the shock wave is launched outward, the fluid elements in the inner region of the core stagnate to form a PNS and those in the outer region of the core keep moving outward.  
It is seen that the radii at which the fluid elements with the present EOS stagnate are slightly smaller than those with the Shen EOS, i.e., the PNS born at the center of the core with the present EOS is more compact, releasing more gravitational energy due to the relative softness of the present EOS.  
Owing to this larger energy release, the fluid elements with the present EOS at larger radii move outward faster than those with the Shen EOS, and, correspondingly, the shock wave propagates faster.  
Finally, the shock waves reach the edges of the iron cores in both cases: 
This occurs 11 ms after the bounce with the variational EOS, which is 1 ms faster than that with the Shen EOS.
This implies that, in both cases, the explosions are successful in this adiabatic approximation.  
In fact, most of the gravitational energy released by the core collapse is converted into the internal energy of the core and the kinetic energy of the fluid elements in the outer region.  
Although some of the energy is consumed by photodisintegration of iron when the shock wave propagates across the outer region of the core, the released energy is large enough to overcome such an energy loss.  
The explosion energy calculated with the definition given in Ref. \cite{KS3} (which calls for subtracting the baryon mass energy of the ejecta from their gravitational mass energy) is $1.7\times10^{51}$ erg, which is larger than that in the case of the Shen EOS: $1.5\times10^{51}$ erg.  

\begin{figure}
  \centering
  \includegraphics[width=10.0cm]{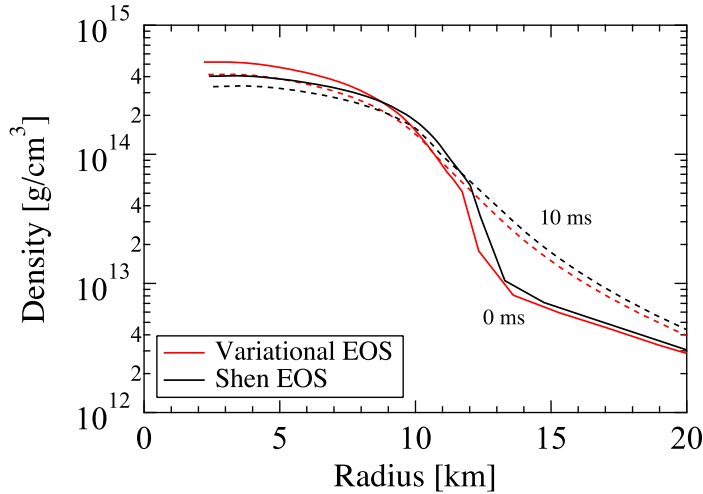}
  \caption{The density profiles at and after the bounce ($t_{\mathrm{pb}}=0$ and 10 ms) as functions of the radius.  
  The results with the Shen EOS are also shown. }
\label{fig:densityTS}
\end{figure}

Figure \ref{fig:densityTS} shows the density profiles in the core at $t_{\mathrm{pb}}= 0$ and 10 ms with both the variational and Shen EOSs.  
The central density with the variational EOS at the bounce is $5.2\times10^{14}$ g/cm$^3$, which is about twice the normal nuclear density, and it can be seen that the central density with the variational EOS at the bounce is higher than that with the Shen EOS.  
Correspondingly, the radius of the high-density core with the variational EOS is smaller.  
Namely, the high-density core region is more compact with the variational EOS, because the variational EOS is softer than the Shen EOS in this density region, as mentioned above. 
Here, we note that the relative softness of the present EOS is consistent with the fact that the incompressibility of symmetric nuclear matter with the variational EOS at the saturation density, 
$K = 245$ MeV, is smaller than that of the Shen EOS, $K = 281$ MeV.  
After the bounce ($t_{\mathrm{pb}}=10$ ms), the central densities slightly decrease and the high-density regions spread for both of the EOSs.  
These results are consistent with the trajectories of the fluid elements in the inner region shown in Fig. \ref{fig:radius}.  
It is noted that the variational EOS used in this simulation is sufficiently causal:
Since the highest density in the present simulation is the central density at the bounce, the matter in the density region used in this simulation is causal, as discussed in the last section.  
This fact is confirmed directly by the analysis of the sound velocity profiles in the present simulation.  

\begin{figure}
  \centering
  \includegraphics[width=10.0cm]{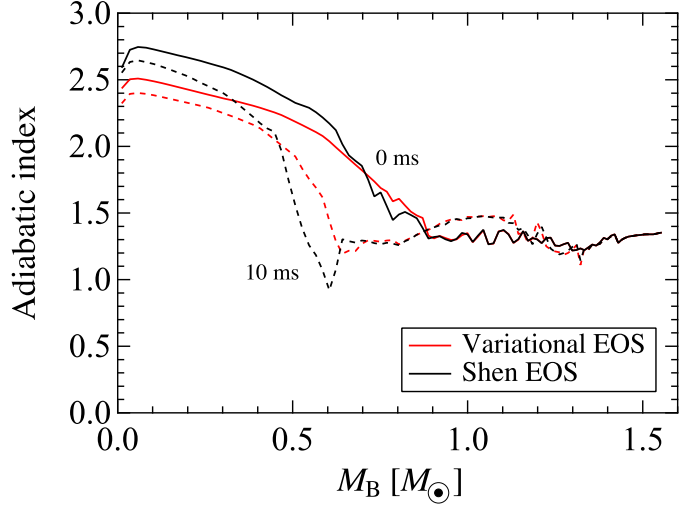}
  \caption{The adiabatic indices at $t_{\mathrm{pb}} = 0$ and 10 ms as functions of the baryon mass coordinate $M_{\mathrm{B}}$.  
  The results with the Shen EOS are also shown. }
\label{fig:gammaTS}
\end{figure}

Figure \ref{fig:gammaTS} shows the adiabatic indices $\Gamma$ at $t_{\mathrm{pb}}=$ 0 and 10 ms as functions of the baryon mass coordinate $M_{\mathrm{B}}$; 
also shown are those with the Shen EOS. 
The figure shows that, at the bounce, the $\Gamma$ in the central region with high densities are significantly larger compared with those in the outer region where $\Gamma \sim$ 4/3, the critical value with respect to gravitational instability.  
In the central region of the core, $\Gamma$ with the present EOS is smaller than that with the Shen EOS, i.e., the present EOS is softer.  
A similar tendency is seen at $t_{\mathrm{pb}} = 10$ ms.  
 
\begin{figure}
  \centering
  \includegraphics[width=10.0cm]{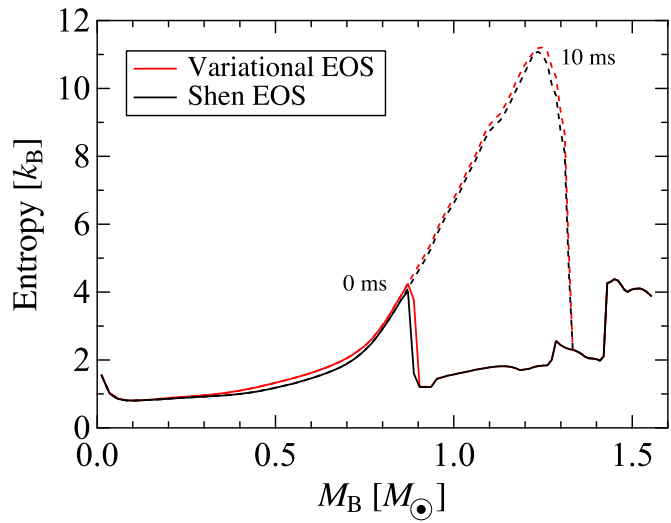}
  \caption{The entropy profiles at $t_{\mathrm{pb}}=0$ and 10 ms as functions of the baryon mass coordinate $M_{\mathrm{B}}$.  
  The results with the Shen EOS are also shown. }
\label{fig:entropyTS}
\end{figure}

Figure \ref{fig:entropyTS} shows the profiles of the entropy $S$ in the core as functions of the baryon mass coordinate $M_{\mathrm{B}}$ at $t_{\mathrm{pb}}=0$ and 10 ms, with our EOS and the Shen EOS. 
At the bounce, the entropy in the inner part of the core remains at about 1--1.5 because of the adiabatic collapse, and, behind the shock, the entropy increases steeply.  
The maximum entropy at the bounce is $S \sim$ 4.  
As the shock wave propagates, the entropy behind the shock increases significantly, and, at $t_{\mathrm{pb}}=10$ ms, $S$ reaches about 11.  
It can also be seen that, at $t_{\mathrm{pb}}=0$ and 10 ms, the $S$ with the present EOS are close to those with the Shen EOS. 

\begin{figure}
  \centering
  \includegraphics[width=10.0cm]{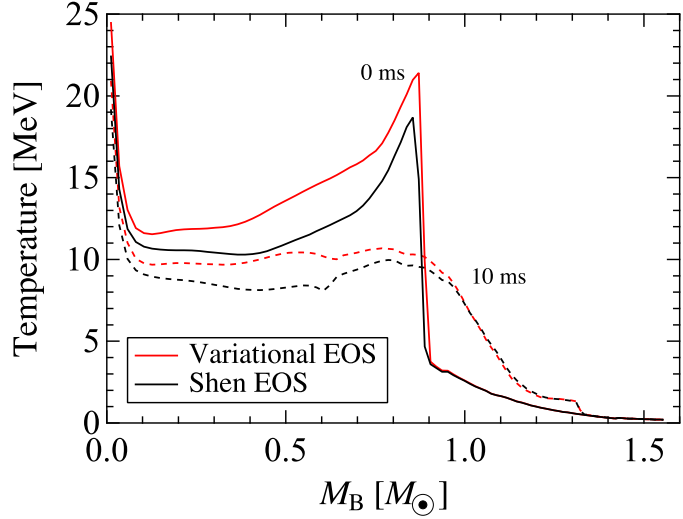}
  \caption{The temperature profiles at $t_{\mathrm{pb}}=0$ and 10 ms as functions of the baryon mass coordinate $M_{\mathrm{B}}$.  
  The results with the Shen EOS are also shown. }
\label{fig:temperatureTS}
\end{figure}

Figure \ref{fig:temperatureTS} shows the temperature profiles in the core as functions of the baryon mass coordinate $M_{\mathrm{B}}$ at $t_{\mathrm{pb}}=0$ and 10 ms.  
It can be seen that, at the bounce, the temperatures $T$ behind the shock are much higher than those in front of the shock for both of the EOSs; in particular, the temperatures $T$ just behind the shock have sharp peaks (with the present EOS, the peak has a height of more than 20 MeV).  
Figure \ref{fig:temperatureTS} also shows that the $T$ in the core with the present EOS are higher than those with the Shen EOS: 
the temperature $T$ just behind the shock with the present EOS is higher by a few MeV.  
After the bounce ($t_{\mathrm{pb}}=10$ ms), the profile behind the shock becomes relatively flat, and $T$ with the present EOS is still higher than that with the Shen EOS.  
These differences between the profiles of $T$ with these two EOSs are due to the relative compactness of the high-density core with the variational EOS: 
Since the present EOS is softer than the Shen EOS, the density of the central core with the present EOS is higher than that with the Shen EOS, and then the higher $T$ follows.  

\begin{figure}
  \centering
  \includegraphics[width=10.0cm]{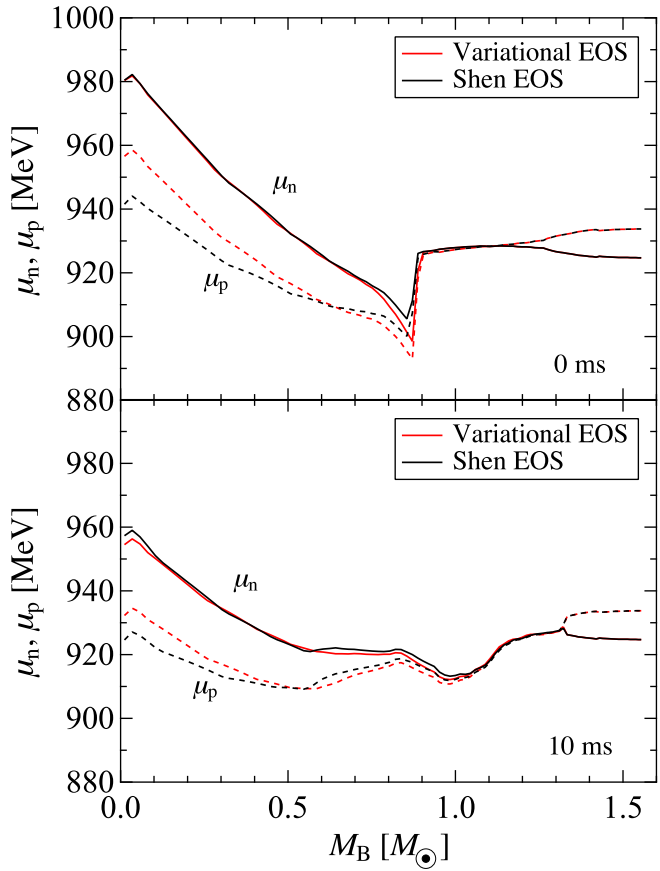}
 \caption{The proton chemical potential $\mu_{\mathrm{p}}$ and the neutron chemical potential $\mu_{\mathrm{n}}$  at $t_{\mathrm{pb}}=0$ ms (upper panel) and 10 ms (lower panel) as functions of the baryon mass coordinate $M_{\mathrm{B}}$.  
The results with the Shen EOS are also shown. }
\label{fig:muTS}
\end{figure}

Figure \ref{fig:muTS} shows the chemical potentials of neutrons $\mu_{\mathrm{n}}$ and protons $\mu_{\mathrm{p}}$ at 
$t_{\mathrm{pb}}=0$ and 10 ms.  
At the bounce, the shock is at $M_{\mathrm{B}} \sim 0.9$ $M_{\odot}$, and $\mu_{\mathrm{p}} \sim \mu_{\mathrm{n}}$ at $M_{\mathrm{B}} \gtrsim 0.9$ $M_{\odot}$ because $Y_{\mathrm{p}} \sim 0.5$ in front of the shock.  
Since the fluid elements just behind the shock are at high $T$, the chemical potentials $\mu_{\mathrm{p}}$ and $\mu_{\mathrm{n}}$ abruptly become lower than those in front of the shock. 
In the inner region, the density increases, and then $\mu_{\mathrm{p}}$ and $\mu_{\mathrm{n}}$ increase.  
At $t_{\mathrm{pb}}=10$ ms, the density profiles become flatter in this region (see Fig. \ref{fig:densityTS}), and, correspondingly, the profiles of $\mu_{\mathrm{n}}$ and $\mu_{\mathrm{p}}$ are also flatter.  
In both cases at $t_{\mathrm{pb}}=0$ and 10 ms, it can be seen that the average of $\mu_{\mathrm{p}}$ and $\mu_{\mathrm{n}}$ with the variational EOS is higher than that with the Shen EOS, because the corresponding $\rho_{\mathrm{B}}$ with the variational EOS is higher.  
Furthermore, the difference between $\mu_{\mathrm{p}}$ and $\mu_{\mathrm{n}}$ with the present EOS is smaller, because, as pointed out in the last section, the symmetry energy $E_{\mathrm{sym}}$ with the variational EOS is smaller than that with the Shen EOS.  
As a result, the difference between $\mu_{\mathrm{n}}$ with the variational and Shen EOSs is smaller than that between $\mu_{\mathrm{p}}$ with the two EOSs, as seen in Fig. \ref{fig:muTS}, though the good agreement between $\mu_{\mathrm{n}}$ with the present and Shen EOSs is rather accidental. 
%%%%%%%%%%%%%%%%%%%%%%%%%%%%%%%%%%%%%%%%%%%%%%%%%%%%%%%%%%%%%%%%%%%%%%%%%%%
\section{Application to core-collapse supernovae with neutrino transfer}
In this section, we take into account the weak reactions, which were neglected in the last section, and apply the EOS constructed in the last section to a spherical SN simulation with neutrino transfer, 
as a more realistic case.  
During the collapse and bounce of the presupernova core and the stage after the bounce, various weak reactions take place; then, a large number of neutrinos created through those reactions escape from the core carrying a large amount of energy away from the collapsed core.  
Due to these neutrino emissions and energy loss, 
the SN explosion is not successful in modern spherically symmetric numerical simulations \cite{KS4, SN1, SN2, SN3}. 
In this case, the core becomes neutron-rich owing to the electron-capture reactions, i.e., the hydrodynamics is governed by the SN-EOS at $Y_{\mathrm{p}}$ values smaller than that of the adiabatic simulation.  
In this section, we adopt the numerical code of the general-relativistic spherically symmetric SN hydrodynamic simulations including neutrino transfer used in Ref. \cite{KS4}, and perform the SN simulation starting from the presupernova model of 15 $M_\odot$ by Woosley and Weaver  \cite{WW} as before.  
Details of the numerical calculations of hydrodynamics and the weak reactions included in this simulation are found in Refs. \cite{KS4, SY2}.  
As in the last section, we are interested in investigating the properties of the variational EOS in the SNe.  
Therefore, we focus on the behavior of the thermodynamic quantities in the high-density core region.  

\begin{figure}
  \centering
  \includegraphics[width=10.0cm]{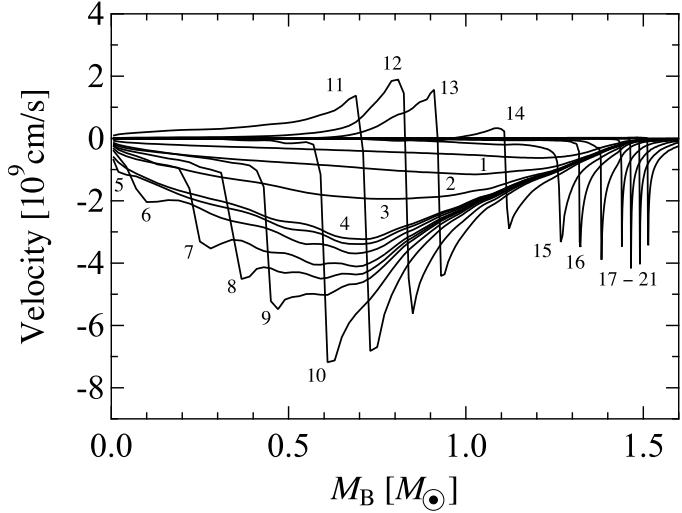}
  \caption{The velocity profiles at selected times as functions of the baryon mass coordinate $M_{\mathrm{B}}$. 
The numbers denote the times explained in the text.} 
\label{fig:velocity2}
\end{figure}

Figure \ref{fig:velocity2} shows the time evolution of the velocity profiles obtained by the simulation including the neutrino transfer with the present variational EOS.  
The numbers denote the time sequence as in Fig. \ref{fig:velocity}: 
The numbers (1)--(4) correspond to the times when the central density reaches $10^{11}$, $10^{12}$, $10^{13}$, and $10^{14}$ g/cm$^3$ during the collapse, respectively. 
The corresponding values of $t_{\mathrm{pb}}$ are $-24.5$, $-5.80$, $-1.54$, and $-0.46$ ms, which are different from those shown in Table \ref{table2}, because the collapses with and without the neutrino transfer are different from each other.  
The numbers (5)--(16) represent the times around and after the bounce, and the values of the corresponding $t_{\mathrm{pb}}$ are the same as those shown in Table \ref{table2}. 
The numbers (17)--(21) correspond to $t_{\mathrm{pb}}=$100, 200, 300, 400, and 500 ms, respectively. 
As in the adiabatic case, a pressure wave appears at the stage corresponding to (5) and it becomes a shock wave at the bounce (10). 
After the bounce, the shock propagates outward, but, in contrast with the adiabatic case, the shock wave stalls at (17)--(21) due to the energy loss by the neutrino emission.  
During the simulation ($t_{\mathrm{pb}} \leq 500$ ms), there is no sign of revival of the shock due to heating by the neutrinos emitted from the inside of the core.  
This result is consistent with other modern 1D SN simulations \cite{KS4, SN1, SN2, SN3}.  

\begin{figure}
  \centering
  \includegraphics[width=10.0cm]{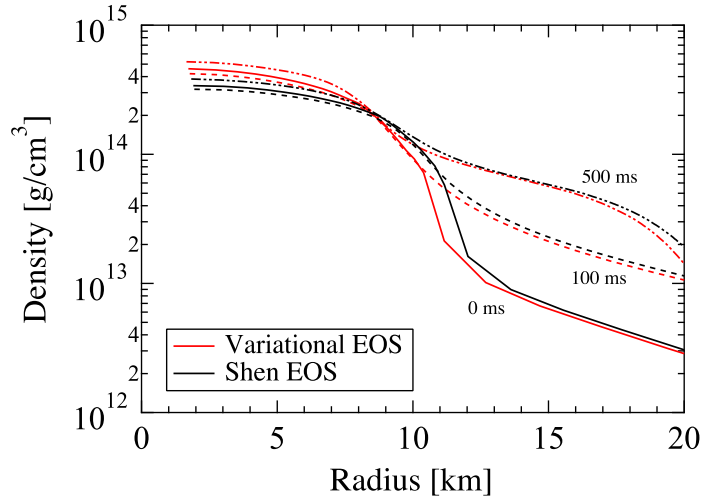}
  \caption{The density profiles at $t_{\mathrm{pb}} =$ 0, 100, and 500 ms as functions of the radius. 
The results with the Shen EOS are also shown. }
\label{fig:density2}
\end{figure}

Figure \ref{fig:density2} shows the density profiles at $t_{\mathrm{pb}} =$ 0, 100, and 500 ms.  
Also shown are the results with the Shen EOS.  
As can be seen in Fig. \ref{fig:density2}, the central density with the variational EOS becomes $4.6\times10^{14}$ g/cm$^3$ at the bounce, which is higher than the central density with the Shen EOS ($3.4\times10^{14}$ g/cm$^3$).  
Correspondingly, the radius of the high-density core is smaller than that with the Shen EOS, as seen in this figure.  
These results imply that, even in the case with neutrino transfer, or with the smaller $Y_{\mathrm{p}}$, as seen below, the present EOS is softer than the Shen EOS.  
This tendency is also seen at later phases.  
After the bounce, at $t_{\mathrm{pb}} = 100$ ms, the central density is lower than that at the bounce, but, at $t_{\mathrm{pb}} = 500$ ms, the central density becomes 
higher ($\rho_{\mathrm{B}} = 5.2\times10^{14}$g/cm$^{3}$): 
the inner core is compressed due to the materials accreting from the outside of the core.  
Here, we note that, as in the adiabatic simulation, the EOS in the density region used in the present simulation ($t_{\mathrm{pb}}\leq 500$ ms)   is sufficiently causal: 
this fact is confirmed by the analysis of the sound velocity profiles in this simulation.  

\begin{figure}
  \centering
  \includegraphics[width=10.0cm]{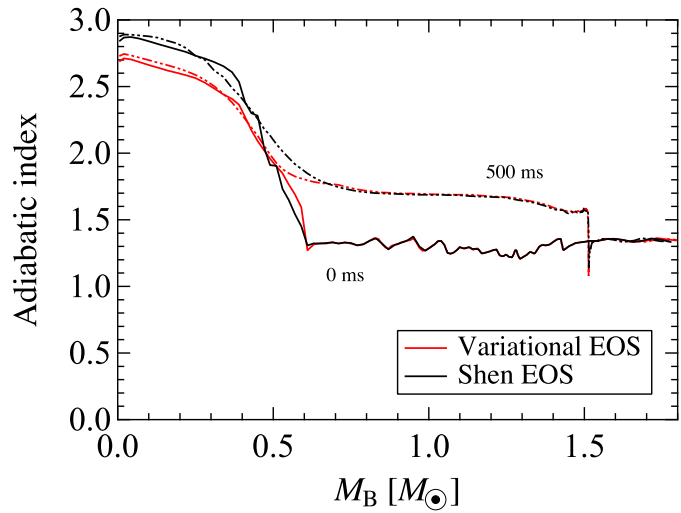}
  \caption{The adiabatic indices $\Gamma$ at $t_{\mathrm{pb}} = 0$ and 500 ms as functions of the baryon mass coordinate $M_{\mathrm{B}}$.  
The results with the Shen EOS are also shown. }
\label{fig:gamma2}
\end{figure}

The adiabatic indices $\Gamma$ at $t_{\mathrm{pb}} = 0$ and 500 ms with the present and Shen EOSs are shown as functions of the baryon mass coordinate $M_{\mathrm{B}}$ in Fig. \ref{fig:gamma2}. 
As in the adiabatic case, at $t_{\mathrm{pb}}=0$ ms, $\Gamma$ in the outer core is close to 4/3, while, in the inner core, $\Gamma$ becomes larger.  
Corresponding to the results obtained from the density profiles, $\Gamma$ in the central region with the variational EOS is smaller than that with the Shen EOS, showing that the variational EOS is softer.  
At $t_{\mathrm{pb}}=500$ ms, $\Gamma$ in the region $M_{\mathrm{B}} \lesssim 1.5$ $M_{\odot}$ is larger than 4/3 because the matter in this region is at high densities, as inferred from Fig. \ref{fig:density2}.  

\begin{figure}
  \centering
  \includegraphics[width=10.0cm]{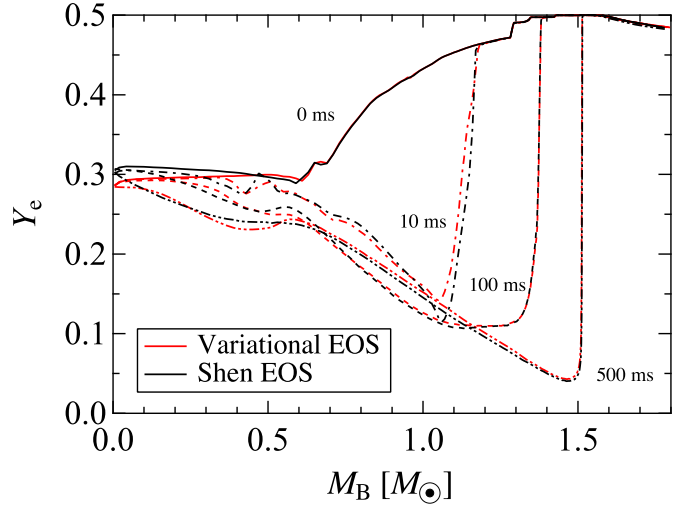}
  \caption{The electron fractions $Y_{\mathrm{e}}$ at $t_{\mathrm{pb}} =$ 0, 10, 100, and 500 ms as functions of the baryon mass coordinate $M_{\mathrm{B}}$.  
  The results with the Shen EOS are also shown. }
\label{fig:ye2}
\end{figure}

Figure \ref{fig:ye2} shows the profiles of the electron fractions $Y_{\mathrm{e}}$ at $t_{\mathrm{pb}} =$ 0, 10, 100, and 500 ms.  
In contrast with the adiabatic case, $Y_{\mathrm{e}}$ varies due to the weak reactions.  
At the bounce ($t_{\mathrm{pb}} = 0$ s), $Y_{\mathrm{e}}$ in the outer-core region decreases toward the center of the core, and, in the inner core ($M_{\mathrm{B}} \lesssim 0.6$ $M_{\odot}$), 
$Y_{\mathrm{e}}$ is about 0.28.  
This value of $Y_{\mathrm{e}}$ is smaller than its initial value ($Y_{\mathrm{e}}=Y_{\mathrm{l}} \gtrsim 0.4$) due to the electron-capture reactions.  
In fact, in the central region of the inner core, the lepton fraction $Y_{\mathrm{l}}$ including the fraction of neutrinos is $Y_{\mathrm{l}} \sim 0.35$, 
which is smaller than the initial value of $Y_{\mathrm{l}}$.  
It is also seen that, at the bounce, $Y_{\mathrm{e}}$ with the present EOS in the central region is smaller than that with the Shen EOS.  
Here, we note that, at the bounce, $Y_{\mathrm{l}}$ with the variational EOS is close to that with the Shen EOS. 
The smaller $Y_{\mathrm{e}}$ at the central region with the present EOS is mainly caused by the smaller symmetry energy $E_{\mathrm{sym}}$, as demonstrated in Fig. \ref{fig:SNMYi}.
At later stages in the evolution, the profiles of $Y_{\mathrm{e}}$ with the present EOS are similar to those in the case with the Shen EOS; $Y_{\mathrm{e}}$ in the outer 
core decreases and, just behind the shock at $t_{\mathrm{pb}} = 500$ ms, $Y_{\mathrm{e}} \sim 0.03$, which exemplifies the necessity for an SN-EOS to cover a wide range of $Y_{\mathrm{e}}$.  
We also note that, at later stages, $Y_{\mathrm{l}}$ decreases toward the surface of the core, as in the case of $Y_{\mathrm{e}}$ shown in this figure.  

\begin{figure}
  \centering
  \includegraphics[width=10.0cm]{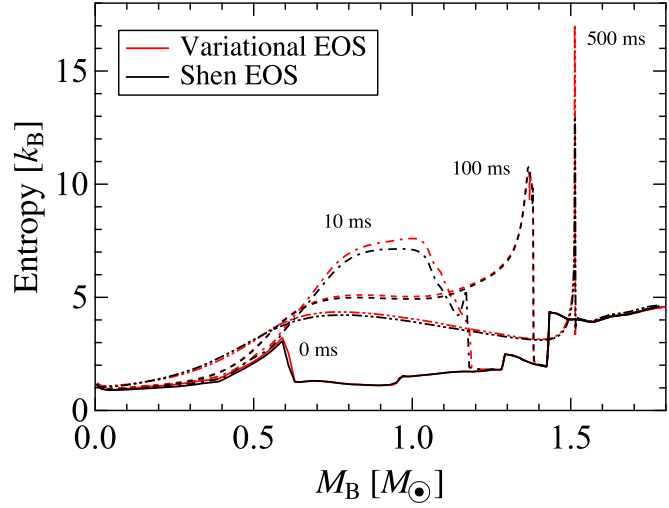}
  \caption{The entropy profiles at $t_{\mathrm{pb}} =$ 0, 10, 100, and 500 ms as functions of the baryon mass coordinate $M_{\mathrm{B}}$.  
The results with the Shen EOS are also shown. }
\label{fig:entropy2}
\end{figure}

Figure \ref{fig:entropy2} shows the profiles of the entropy $S$ at $t_{\mathrm{pb}} =$ 0, 10, 100, and 500 ms with the present and Shen EOSs.  
As in the case of the adiabatic collapse, $S$ has a jump behind the shock (about 3 at the bounce), and $S$ increases, especially behind the shock front as time passes.
In later stages, for $t_{\mathrm{pb}} \gtrsim 100$ ms, the profile of $S$ has a sharp peak just behind the shock and a hump at $M_{\mathrm{B}} \sim 0.8$ $M_{\odot}$.  
$S$ decreases gradually at the outer region behind the shock: 
This negative gradient of $S$, together with the negative gradient of $Y_{\mathrm{l}}$ mentioned above, 
implies the possibility of instability with respect to convection, as discussed in Ref. \cite{KS3}.  
$S$ in the inner region increases gradually, while the hump decreases at later stages. 
It can also be seen that the $S$ with the variational EOS are very close to those with the Shen EOS.  

\begin{figure}
  \centering
  \includegraphics[width=10.0cm]{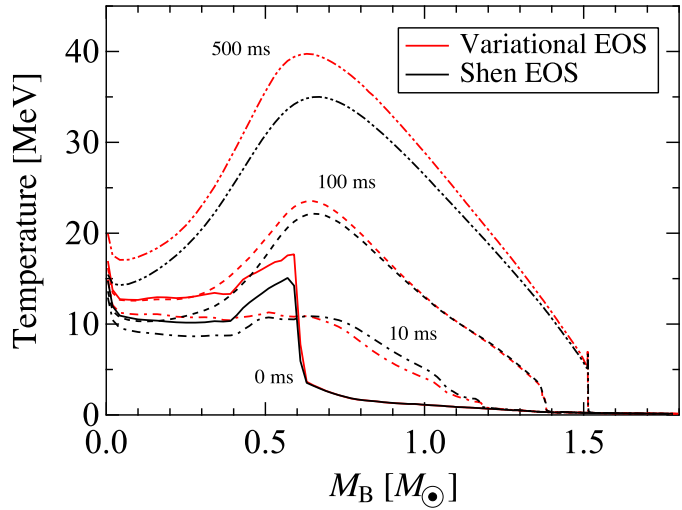}
  \caption{The temperature profiles at $t_{\mathrm{pb}} =$ 0, 10, 100, and 500 ms as functions of the baryon mass coordinate $M_{\mathrm{B}}$.  
The results with the Shen EOS are also shown. }
\label{fig:temp2}
\end{figure}

The temperature profiles are shown in Fig. \ref{fig:temp2}. 
The profile at the bounce is similar to that of the adiabatic calculation: 
At $t_{\mathrm{pb}}=0$ ms, $T$ behind the shock is about 15 MeV with the present EOS, which is higher than that with the Shen EOS, due to the compactness of the inner core with the present EOS.  
Even at later stages, $T$ is still higher with the variational EOS: 
The maximum value of $T$ is about 40 MeV at $t_{\mathrm{pb}}=500$ ms, which is about 5 MeV higher than that with the Shen EOS.  

\begin{figure}
  \centering
  \includegraphics[width=10.0cm]{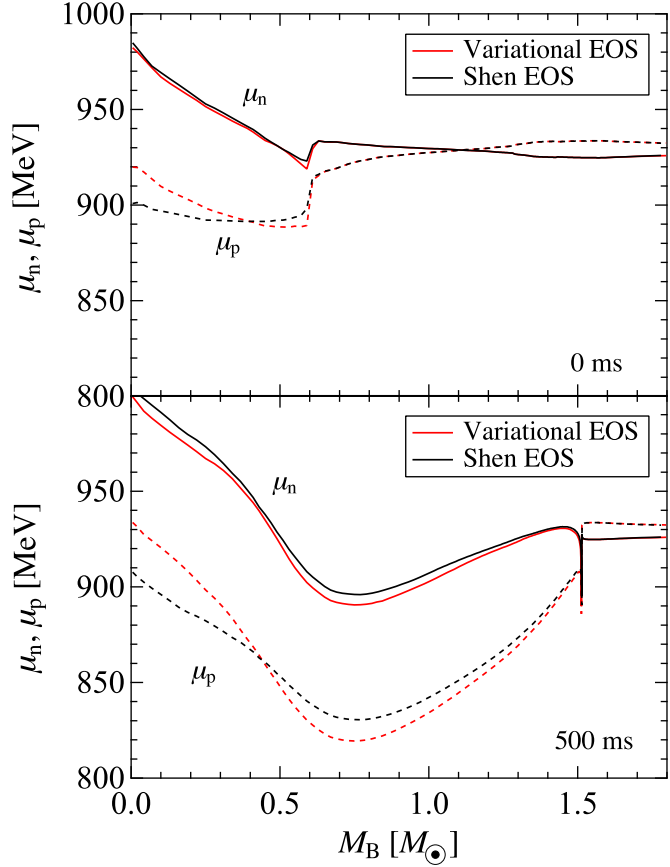}
 \caption{The proton chemical potential $\mu_{\mathrm{p}}$ and the neutron chemical potential $\mu_{\mathrm{n}}$ 
 at $t_{\mathrm{pb}}=0$ ms (upper panel) and 500 ms (lower panel) as functions of the baryon mass coordinate $M_{\mathrm{B}}$.  
 The results with the Shen EOS are also shown. }
\label{fig:mu2}
\end{figure}

Figure \ref{fig:mu2} shows the proton chemical potentials $\mu_{\mathrm{p}}$ and the neutron chemical potentials $\mu_{\mathrm{n}}$ at $t_{\mathrm{pb}}=0$ ms (upper panel) and 500 ms (lower panel).  
It can be seen that, as in the adiabatic case, the average of $\mu_{\mathrm{n}}$ and $\mu_{\mathrm{p}}$ with the variational EOS is higher than that with the Shen EOS due to the higher densities with the present EOS, and, owing to the smaller $E_{\mathrm{sym}}$, the difference between $\mu_{\mathrm{n}}$ and $\mu_{\mathrm{p}}$ is smaller with the variational EOS.  
In fact, $\mu_{\mathrm{n}}$ with the present EOS tends to be close to that with the Shen EOS, while $\mu_{\mathrm{p}}$ tends to be higher in the central high-density region. 
The good agreement between $\mu_{\mathrm{n}}$ with the present and Shen EOSs is rather accidental. 
%%%%%%%%%%%%%%%%%%%%%%%%%%%%%%%%%%%%%%%%%%%%%%%%%%%%%%%%%%%%%%%%%%%%%%%%%%%
\section{Summary and concluding remarks}
In this paper, we have applied the EOS of hot uniform nuclear matter based on the cluster variational method with the AV18 two-body potential and the UIX three-body potential to spherically symmetric SN numerical simulations. 
To our knowledge, this is the first application of the many-body nuclear EOS based on bare nuclear forces to SN numerical simulations.  
Before the application to SN simulations, we examined uniform isentropic SN matter with fixed lepton number fractions. 
The temperatures of isentropic SN matter ($S =1$ and 2) with the variational EOS were close to  but slightly lower than those with the Shen EOS. 
The SN matter with the variational EOS was more neutron-rich, because the symmetry energy of the variational EOS, $E_{\mathrm{sym}}$ = 30.0 MeV, is smaller than that of the Shen EOS, $E_{\mathrm{sym}}$ = 36.9 MeV.  
It was also found that the pressures with the variational EOS are lower than those with the Shen EOS.  

In the application of the variational EOS to SN simulations, it was supplemented by the Shen EOS for low-density non-uniform matter.  
Starting from the 15 $M_{\odot}$ presupernova model constructed by Woosley and Weaver, we performed the general-relativistic spherically symmetric SN simulations without neutrino transfer, as a test calculation, and with neutrino transfer, as a more realistic case.  
We found that the numerical codes run successfully with the present EOS, i.e., the present EOS is applicable to more realistic SN numerical simulations.  
In the adiabatic calculation without the weak reactions, it was seen that the stellar core created after the bounce with the variational EOS is more compact than that with the Shen EOS, implying that the present EOS in the adiabatic case ($Y_{\mathrm{p}} \gtrsim 0.4$) is softer than the Shen EOS.  
This softness of the present EOS results in a larger explosion energy of $1.7\times10^{51}$ erg, as compared with the case of the Shen EOS.  
Here, we note that this relative softness of the variational EOS is consistent with the fact that the incompressibility of the present EOS, $K = 245$ MeV, 
is smaller than that of the Shen EOS, $K = 281$ MeV.  
In the case including neutrino transfer, the shock created after the bounce stalls due to the energy loss by the neutrino emissions.  
Also, in the case with smaller $Y_{\mathrm{p}}$, it was seen that the present EOS is softer than the Shen EOS.  
The electron fraction $Y_{\mathrm{e}}$ in the central region of the core with the variational EOS is found to be 
smaller than that with the Shen EOS at the bounce. 
This is because $Y_{\mathrm{l}}$ with the present EOS is similar to that with the Shen EOS at the bounce and the symmetry energy with the present EOS is smaller.  

As mentioned before, in the SN simulation with neutrino transfer, we have focused on the thermodynamic properties of the inner core, because our aim is to examine the uniform variational EOS for small values of $Y_{\mathrm{p}}$.  
Therefore, we have not discussed the behavior of the emitted neutrinos, such as their energy spectra or luminosities: 
These quantities are largely affected by the compositions of the SN matter in the non-uniform region where we use the Shen EOS in the present calculations.  

Since the present EOS is consistent with the observational data on cold neutron stars (both the masses and radii) \cite{TT} and provides larger explosion energy than the Shen EOS, 
it has preferable features as an SN-EOS; 
it is certainly desirable to apply this EOS to multi-dimensional SN simulations that are more realistic. 
The application of the variational EOS to 
SN simulations for a series of progenitors with various masses, including one with an O--Ne--Mg core, is also interesting. 

Furthermore, it is fascinating to study the influence of the uncertainty in the three-body force on the SN dynamics.  
In fact, the uncertainty in the three-body force, 
which is deeply related to the symmetry energy, has been considered in studies on neutron stars.  
For example, in Ref. \cite{Gandolfi}, the UIX and the Illinois models of the three-body potentials \cite{Illinois} are employed in Monte Carlo calculations of the neutron matter EOS at zero temperature, and the neutron-star structure with those three-body forces is studied systematically; 
extensions of the present study will enable us to perform similar investigations on the dynamics of the core-collapse SNe.  

Before further advancing those applications and extensions, however, an EOS of non-uniform matter self-consistently based on the variational EOS must inevitably be constructed.  
In fact, the EOS of non-uniform matter determines the initial condition, electron capture rates, 
dynamics of the collapsing core, and various neutrino reactions at later stages after bounce, in the SN simulations. 
Construction of the non-uniform EOS that is consistent with the present uniform EOS is now in progress, and will be reported elsewhere in the near future.  
%%%%%%%%%%%%%%%%%%%%%%%%%%%%%%%%%%%%%%%%%%%%%%%%%%%%%%%%%%%%%%%%%%%%%%%%%%%

\section*{Acknowledgments}
We would like to express special thanks to H. Kanzawa, H. Suzuki, M. Yamada, and S. Yamada for valuable discussions and comments, and to H. Matsufuru for support with parallel computing. 
The numerical computations in this work were carried out on SR16000 at YITP in Kyoto University, on the supercomputer (NEC SX8R) at the Research Center for Nuclear Physics, Osaka University, on the GPGPU and SR16000 at the High Energy Accelerator Research Organization (KEK), and on SR16000 at the Information Technology Center of the University of Tokyo.
This work is supported by JSPS (Nos. 21540280, 22540296, 24244036, 25400275), the Grant-in-Aid for JSPS Fellows (No. 24-3275), 
and the Grant-in-Aid for Scientific Research on Innovative Areas of MEXT (Nos. 20105003, 20105004, 20105005, 24105008).  


%\bibliographystyle{ptephy}
%\bibliography{sample}
%
% once the .bbl file has been generated then place the text in your article.

\vfill\pagebreak

\end{document}